\newcommand{\apj}{Astrophys. J.}             
\newcommand{\aap}{Astron. and Astrophys.}
\newcommand{\apjl}{Astrophys. J. Letters}
\documentclass{iaus}
\usepackage{graphicx}

\title[Outskirts of Galaxy Clusters: intense life in the suburbs]{
Warm gas in the outskirts of galaxy clusters -- The cluster soft excess phenomenon}

\author{Massimiliano~Bonamente$\,^{1,2}$, 
Richard~Lieu$\,^{1}$ and Jonathan~P.D.~Mittaz$\,^{1}$
}

\affiliation{\(^{\scriptstyle 1} \){Department of Physics, University of Alabama in Hunstville,
Huntsville, AL}\\
\(^{\scriptstyle 2} \){National Space Science and Technology Center, NASA/MSFC, Huntsville, AL}\\
}

\pubyear{2004}
\volume{195}
\pagerange{1--8}
\date{?? and in revised form ??}
\jname{Outskirts of Galaxy Clusters: intense life in the suburbs}
\editors{A. Diaferio, ed.}
\begin{document}
                                                                                                                                                                                                
\maketitle

\begin{abstract}
The cluster soft excess emission indicates the presence of large 
amounts of warm gas (T$\sim10^6$ K) in the neighborhood of galaxy
clusters. Among the clusters that display this phenomenon is the Coma cluster, the nearest rich galaxy cluster.
The excess emission is more prominent at the cluster's outskirts
than at its center.
Detailed studies of its large-scale emission -- up to $\sim$2.6 Mpc from the cluster's center --
reveal that these warm baryons are as massive as, or
possibly more massive than, the well-known hot intra-cluster medium (T$\sim10^8$ K). 
A possible interpretation of the excess emission from the Coma cluster is radiation from low-density filaments
located in the neighborhood of the cluster. In this case, the filaments would extend for much larger distances,
or feature higher density, than predicted by current cosmological simulations.

\end{abstract}

\section{Introduction to the soft excess emission}

Clusters of galaxies are strong emitters of X-rays, which originate from a hot and
diffuse intra-cluster medium (ICM) at temperatures of a few $\times 10^{7}$ K.
The soft X-ray band below $\sim$ 1 keV often feature the `soft excess' emission phenomenon, i.e., 
radiation in excess of that expected from the hot ICM.
The excess emission could originate from Inverse-Compton scattering of cosmic microwave background (CMB) photons
against a population of relativistic electrons in the intra-cluster medium 
(Hwang 1997, Sarazin and Lieu 1998, Ensslin and Biermann 1998, Lieu et al. 1999). 
Alternatively, warm gas at T$\sim 10^6$ K could be responsible for the soft emission
(e.g., Lieu et al. 1996a,b, Nevalainen et al. 2003, Kaastra et al. 2003, Bonamente, Joy and Lieu 2003).
Warm gas may reside inside the cluster, or in very diffuse filamentary structures
outside the cluster, as seen in
large scale hydrodynamic simulations (e.g., Cen and Ostriker 1999, Dav\'{e} et al. 2001, Cen et al 2001).
The warm gas scenario appears to be favored by the
current X-ray data (e.g., Bonamente, Lieu and Mittaz 2001a, Buote 2001, Kaastra et al. 2003) 

\section{Spatial and spectral features of the soft excess emission}

The soft excess emission is  present in $\sim$ 50\% of the clusters.
Two surveys by Bonamente et al. (2002) and Kaastra et al. (2003)
reported, respectively, detection of soft excess emission in 18/38 clusters and
in 7/21 clusters that were investigated. Also, the excess is normally detected when the observations
have a high S/N, such as in the nearby Virgo (Bonamente, Lieu and Mittaz 2001a)
and Coma clusters (Bonamente, Joy and Lieu 2003).

The relative importance of the soft excess component -- with respect to the hot  ICM component --
normally increases with radial distance from the cluster center. This is the case of
the Coma cluster (Fig.1; Bonamente, Joy and Lieu 2003), of A1795 (Mittaz et al. 1998),
A2199 (Lieu, Bonamente and Mittaz 1999), Virgo (Bonamente, Lieu and Mittaz 2001a;
Lieu et al. 1996a), MKW3s and A2052 (Kaastra et al. 2003).
An important exception in AS1101 (Bonamente, Lieu and Mittaz 2001a), where the soft
excess is sharply peaked at the cluster's center.

\begin{figure}
\includegraphics[angle=90,width=5.5in]{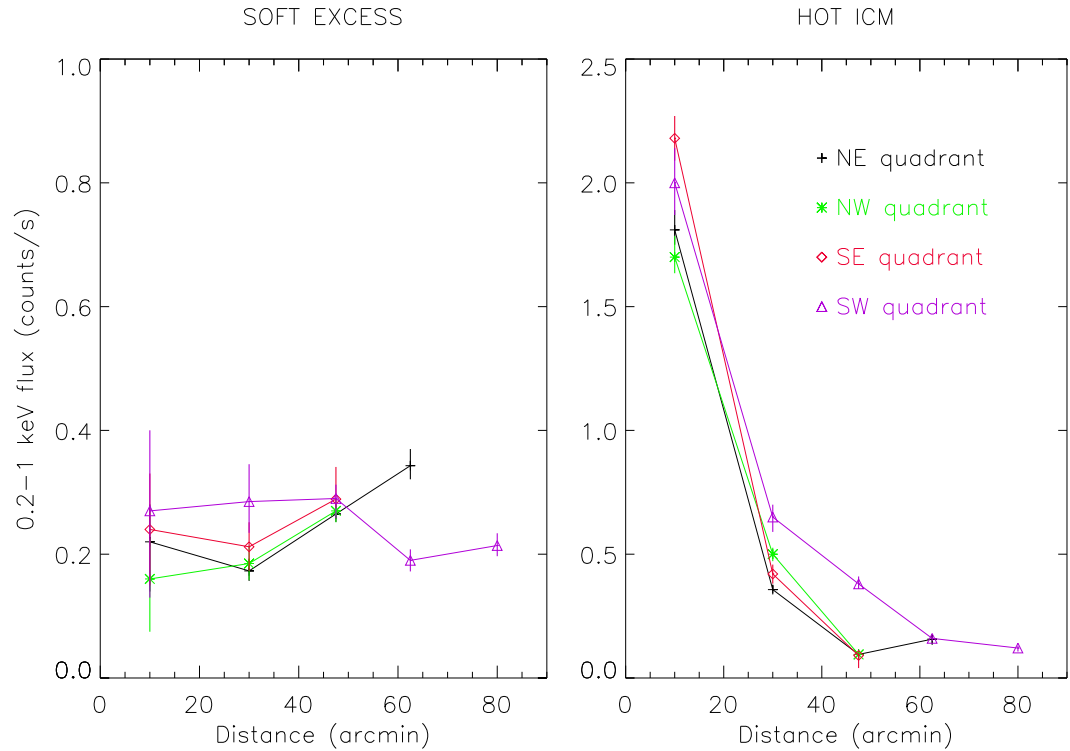}
\caption{\scriptsize The radial distribution of the soft excess emission 
and of the hot ICM emission of Coma from
ROSAT PSPC data. The data was analyzed in concentric annuli, further divided into quadrants.}
\vspace{-0.0cm}
\end{figure}
                                                                                                            
Thermal modelling of the soft excess component is normally favored by the
goodness-of-fit analysis. Among the highest S/N data analyzed to date,
the Coma cluster PSPC data favors the thermal interpretation for the majority 
of its regions (Fig. 2; Bonamente, Joy and Lieu 2003). The detection of OVII emission lines
in the soft X-ray spectra of a few clusters, reported by Kaastra et al. (2003) 
from XMM data, would be the unmistakable signature of the thermal nature of the soft excess.
A confirmation of those lines with higher S/N is however required before the detection
be regarded as definitive.

The non-thermal interpretation remains a plausible explaination for the excess emission, and it cannot be
formally rejected in its entirety. 
In order
to explain the detected excess emission as non-thermal radiation, some clusters require that the
cosmic rays be near or above pressure equipartition with the hot gas (Lieu et al. 1999; Bonamente,
Lieu and Mittaz 2001a).  

\begin{figure}
\includegraphics[angle=90,width=5.5in]{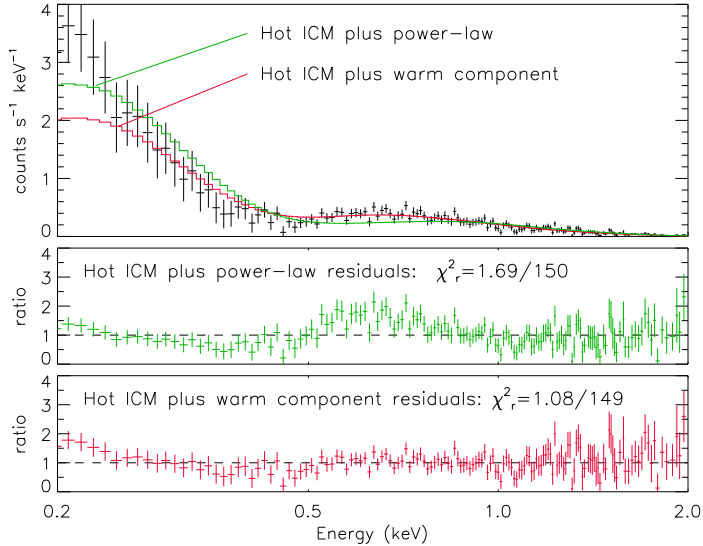}
\vspace{-2.7cm}
\caption{\scriptsize {\it ROSAT} PSPC spectrum of the 50-70' north-eastern quadrant of Coma.
In green is the hot ICM
 plus power law model, in red the hot ICM model plus a low-energy thermal component.}
\vspace{-0cm}
\end{figure}

\section{Thermal interpretation of the soft excess emission}

If the excess emission is of thermal origin, it is possible to envisage two scenarios:\\
(1) The `warm' gas coexists with the `hot' gas. In this case, it is possible to estimate the
density of the warm gas from the emission integral $I=\int n^2 dV$; in Coma
the warm gas would have a density of 9$\times 10^{-4}$ cm$^{-3}$
to $\sim 8 \times 10^{-5}$ cm$^{-3}$ (Bonamente, Joy and Lieu 2003).

If the soft excess emission originates from a warm phase of the intra-cluster medium,
 the ratio of the emission integral ($I=\int n^2 dV$) of the hot ICM
and  the  emission integral of the warm gas
can be used to measure the relative mass of the two phases.
The emission integral is readily measured by fitting the X-ray spectrum.
For the PSPC data of the
Coma cluster, we calculated
$M_{warm}/M_{hot}$=0.75 within a radius of 2.6 Mpc (Bonamente, Joy and Lieu 2003).

(2) The `warm' gas resides in
low density filamentary structures outside the clusters, the warm-hot intergalactic medium (WHIM).
This scenario follows several cosmological simulations (e.g., Cen and Ostriker 1999) which identify
the majority of the low-redshift baryons in a tenuous network of WHIM filaments.
Following this interpretation, the gas should feature overdensities 
of $\delta \sim 3-300$, corresponding to 
$\sim 10^{-4}-10^{-6}$ cm$^{-3}$ (for $\Omega_b \sim 0.05$), and a median overdensity of $\sim 10-30$ (Dav\'{e} et al. 2001). 
We employ a simple model where constant-density filaments are directed towards the observer.
In this case, the emission integral of the soft phase becomes $I= n^2 A L$, where $A$ is the projected area of the filaments
and $L$ their length along the line of sight. In order to explain the detected $I$, filaments should
extend for several Mpc in front of the cluster (see Table 1). Consider the case of filaments with 
density $n=10^{-5}$ cm$^{-3}$  ($\delta \sim 30$): the filaments should then extend for several hundred mega-parsecs!
This figure is at odds with typical results from cosmological simulations, and with the fact that the Coma
cluster is located at a distance of $\sim$ 95 Mpc from the Galaxy. Filaments of higher density ($\sim 10^{-4}$ cm$^{-3}$)
will require shorter $L$ (Table 1); in this case the scenario becomes tenable (Bonamente, Joy and Lieu 2003), although
still at odds with the results from cosmological simulations, which predict an average filament density considerably lower than
$10^{-4}$ cm$^{-3}$.
This result is consistent with a detailed analysis of the soft X-ray emission predicted by the WHIM filaments
seen in some recent simulations (e.g., Cen et al. 2001). The analysis reveals that the WHIM filaments
predict several times ($\geq 10$) lower soft X-ray fluxes than those of the typical 
soft excess emission (Mittaz et al. 2004) as, e.g., that of the Coma cluster.
Similar results apply to the soft excess of the AS1101 cluster, where
a strong centrally-peaked soft excess emission was detected (Bonamente, Lieu and Mittaz 2001a).

\begin{table}
\begin{center}
\caption{Lenght of filaments for the Coma PSPC soft excess emission}
\begin{tabular}{lccc}
\\
Region  & $I^{*}$   &  $L$ (Mpc) & $L$ (Mpc) \\
(arcmin)    &       & [$n=10^{-5}$ cm$^{-3}$] & [$n=10^{-4}$ cm$^{-3}$] \\
\\
0-20  & 0.057 & 1,800 & 18 \\
20-40 & 0.039 & 350 & 3.5 \\
40-55 & 0.066 & 610 & 6.1 \\
55-70 & 0.019 & 250 & 2.5 \\
70-90 & 0.007 & 220 & 2.2 \\
\\
\end{tabular}
 \end{center}
{\scriptsize (*) The emission integral $I$ is in units of $10^{14}[ 4 \pi (1+z)^2
D^2]$,  as in the XSPEC optically-thin MEKAL code. $D$ is distance to the Coma
cluster in cm and $z$=0.023 the redshift. Description of the PSPC  data used to
obtain the emission integrals $I$ can be found in Bonamente, Joy and Lieu (2003).}
\end{table}

Following this scenario,  and for a filament density of $n=10^{-4}$ cm$^{-3}$,
the PSPC data of Coma  yields the conclusion that $M_{fil}/M_{hot}$=3 within
a radius of 2.6 Mpc (Bonamente, Joy and Lieu 2003). The warm gas would therefore be 
more massive than the hot ICM if it is distributed in low-density
filaments.

As the total mass of the Coma cluster within 14 Mpc is 1.6$\pm0.4 \times 10^{15} M_{\bigodot}$
(Geller, Diaferio and Kurtz 1999) and  the mass of the  hot ICM is
$\sim 4.3 \times 10^{14} M_{\bigodot}$ within 2.6 Mpc (Mohr, Mathiesen and Evrard 1999),
the soft excess emission could account for a significant fraction of
the low-redshift $\Omega_b$.
 Kaastra et al. (2003) resported similar results
from XMM observations of several clusters including AS1101, MKW3s and A2052.

\section{Conclusions}
Detection of soft X-ray excess emission from galaxy clusters is commonplace. The radiation
is more prominent at the cluster outskirts, and it indicates that clusters
may be a significant reservoir of `warm' baryons.
Precise mass estimates for the `soft' component require knowledge of the exact location
of the emitter. If the gas is located in filaments outside the clusters, 
the detection of soft excess in the Coma cluster indicates that the `warm' gas is
more massive than the hot ICM. Filaments of considerably higher density than predicted by
the current cosmological simulations are needed to explain the detected excess emission
from the Coma cluster.

\end{document}